\documentclass{article}

\PassOptionsToPackage{numbers, compress}{natbib}

\usepackage[final]{neurips_2022}

\usepackage[utf8]{inputenc} %
\usepackage[T1]{fontenc}    %
\usepackage{url}            %
\usepackage{amsfonts}       %
\usepackage{nicefrac}       %
\usepackage{microtype}      %

\usepackage{pifont}

\usepackage{booktabs}
\usepackage{multirow}
\usepackage{graphicx}
\usepackage{float}
\usepackage{mathtools}
\usepackage{ifthen}
\usepackage{bm}
\usepackage{fp}
\usepackage{siunitx}
\usepackage{amsthm}
\usepackage{caption}
\usepackage{subcaption}
\usepackage{xspace}
\usepackage{xcolor}
\usepackage{amsmath, amsthm, amssymb}

\usepackage{wrapfig}
\usepackage{pifont}

\usepackage{algorithm}
\usepackage{algorithmicx}
\usepackage[noend]{algpseudocode}
\usepackage{comment}
\algnewcommand\algorithmicinput{\textbf{Input:}}
\algnewcommand\Input{\item[\algorithmicinput]}
\algnewcommand\algorithmicoutput{\textbf{Output:}}
\algnewcommand\Output{\item[\algorithmicoutput]}
\algnewcommand{\LineComment}[1]{\State \(\triangleright\) #1}

\usepackage[hidelinks,breaklinks,colorlinks]{hyperref}
\hypersetup{
  linkcolor={blue!70!black},
  citecolor={red!70!black},
  urlcolor={blue!70!black}
}

\usepackage{hyperref}

\newcommand{\sys}{\mbox{\textsc{FeatureRE}}\xspace}

\def\Snospace~{\S{}}

\usepackage{color}

\usepackage{soul}
\usepackage{xcolor}

\usepackage[super]{nth}

\title{Rethinking the Reverse-engineering of Trojan Triggers}

\author{%
  Zhenting Wang \\
  Rutgers University \\
  \texttt{zhenting.wang@rutgers.edu}
  \And
  Kai Mei \\
  Rutgers University \\
  \texttt{kai.mei@rutgers.edu} \\
  \AND
  Hailun Ding \\
  Rutgers University \\
  \texttt{hailun.ding@rutgers.edu} \\
  \And
  Juan Zhai \\
  Rutgers University \\
  \texttt{juan.zhai@rutgers.edu} \\
  \And
  Shiqing Ma \\
  Rutgers University \\
  \texttt{sm2283@rutgers.edu} \\
}

\begin{document}

\maketitle

\begin{abstract}

	Deep Neural Networks are vulnerable to Trojan (or backdoor) attacks.
	Reverse-engineering methods can reconstruct the trigger and thus identify affected models.
	Existing reverse-engineering methods only consider input space constraints, e.g., trigger size in the input space.
	Expressly, they assume the triggers are static patterns in the input space and fail to detect models with feature space triggers such as image style transformations.
	We observe that both input-space and feature-space Trojans are associated with feature space hyperplanes.
	Based on this observation, we design a novel reverse-engineering method that exploits the feature space constraint to reverse-engineer Trojan triggers.
	Results on four datasets and seven different attacks demonstrate that our solution effectively defends both input-space and feature-space Trojans.
	It outperforms state-of-the-art reverse-engineering methods and other types of defenses in both Trojaned model detection and mitigation tasks.
	On average, the detection accuracy of our method is 93\%. For Trojan mitigation, our method can reduce the ASR (attack success rate) to only 0.26\% with the BA (benign accuracy) remaining nearly unchanged.
	Our code can be found at \url{https://github.com/RU-System-Software-and-Security/FeatureRE}.

\end{abstract}
\section{Introduction}\label{sec:intro}

DNNs are vulnerable to Trojan attacks~\cite{gu2017badnets, liu2017trojaning,cheng2020deep, doan2021lira,salem2022dynamic,wang2022bppattack}.
After injecting a Trojan into the DNN model, the adversary can manipulate the model prediction by adding a \textit{Trojan trigger} to get the target label.
The adversary can inject the Trojan by performing the poisoning attack or supply chain attack.
In the poisoning attack, the adversary can control the training dataset and injects the Trojan by adding samples with the Trojan trigger labeled as the target label.
In the supply chain attack, the adversary can replace a benign model with a Trojaned model by performing the supply chain attack.
The Trojan trigger is becoming more and more stealthy.
Earlier works use static patterns, e.g., a yellow pad as the trigger, which is known as the input space triggers.
Researchers recently proposed using more dynamic and input-aware techniques to generate stealthy triggers that mix with benign features, which are referred to as the feature space triggers.
For example, the trigger of the feature-space Trojans can be a warping process~\cite{nguyen2021wanet} or a generative model~\cite{cheng2020deep,nguyen2020input,li2021invisible}.
The Trojan attack is a prominent threat to the trustworthiness of DNN models, especially in security-critical applications, such as autonomous driving~\cite{gu2017badnets}, malware classification~\cite{severi2021explanation}, and face recognition~\cite{sarkar2020facehack}.

Prior works have proposed several ways to defend against Trojan attacks, such as removing poisons in training~\cite{du2019robust,chen2018detecting,tran2018spectral}, detecting Trojan samples at runtime~\cite{gao2019strip,chou2018sentinet,doan2020februus,ma2019nic}, etc. 
Many of above methods can only work for one type of Trojan attack.
For example, training and pre-training time defense (e.g., removing poisoning data, training a benign model with poisoning data) fail to defend against the supply chain attack.
Trigger reverse-engineering~\cite{wang2019neural,liu2019abs,shen2021backdoor,guo2019tabor,chen2019deepinspect} is a general method to defend against different Trojan attacks under different threat models.
It works by searching for if there exists an input pattern that can be used as a trigger in the given model.
If we can find such a trigger, the model has a corresponding Trojan and is marked as malicious and vice versa.
Existing reverse-engineering methods assume that the Trojan triggers are static patterns in the input space and develop an optimization problem that looks for an input pattern that can be used as the trigger.
This assumption is valid for input space attacks~\cite{gu2017badnets,liu2017trojaning,chen2017targeted} that use static triggers (e.g., a colored patch).
Feature space attacks~\cite{cheng2020deep,doan2021lira,salem2022dynamic,nguyen2021wanet,nguyen2020input,li2021invisible,lin2020composite} break this assumption.
Existing trigger reverse-engineering methods~\cite{wang2019neural,liu2019abs,shen2021backdoor,guo2019tabor,chen2019deepinspect} constrain the optimization by using heuristics or empirical observations on existing attacks, such as pixel values are in range \([0, 255]\), and the trigger's size is small.
Such heuristics are also invalid for feature space triggers that change all pixels in images.
Reverse-engineering the feature space is challenging.
Unlike input space, there are no constraints that can be directly used.

\begin{wrapfigure}{R}{0.6\textwidth}
    \centering
    \includegraphics[width=0.6\columnwidth]{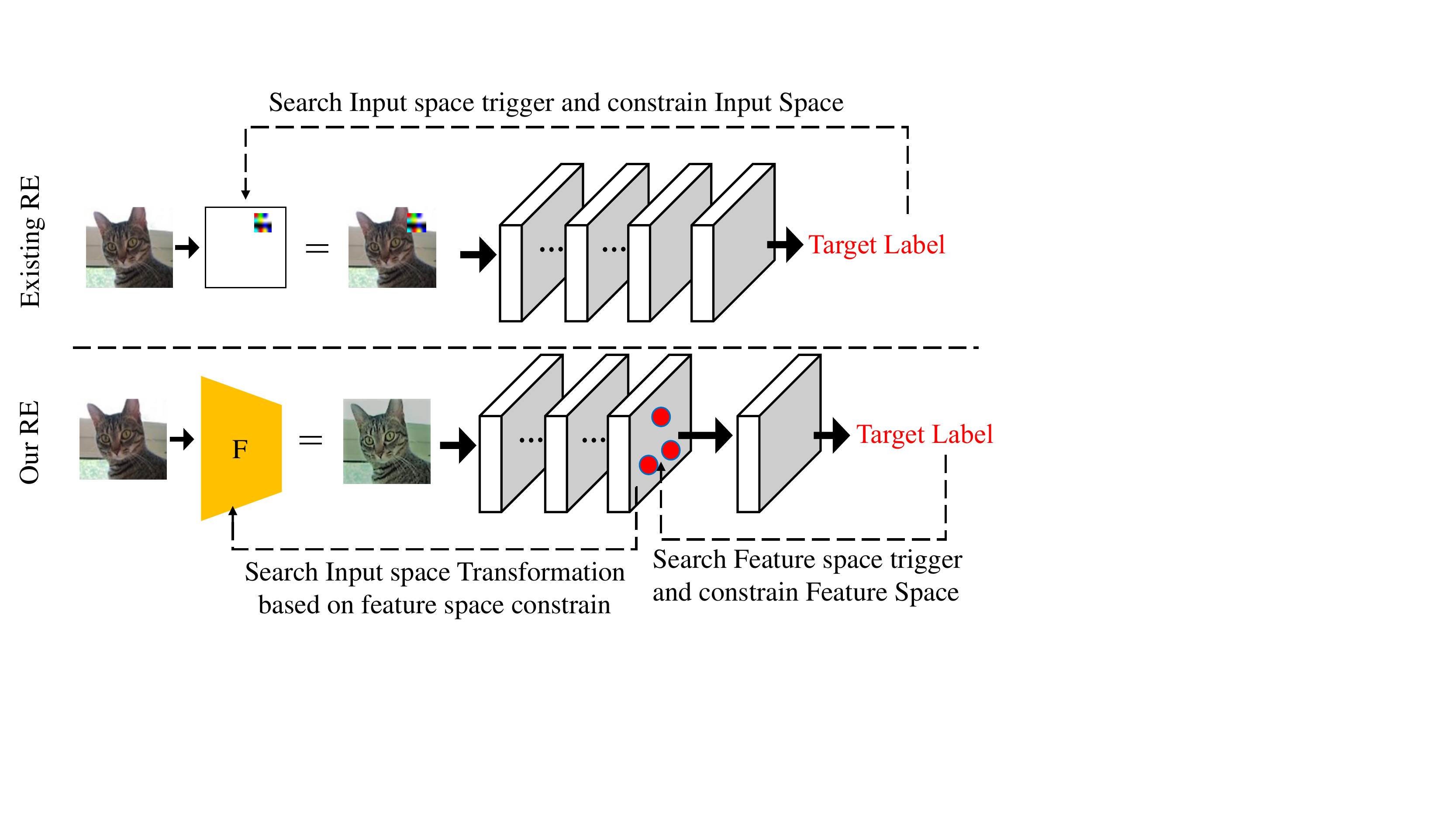}
    \caption{Existing reverse-engineering (RE) and ours.}\label{fig:hyperplane_in_feature}
\end{wrapfigure}

In this paper, we propose a trigger reverse-engineering method that works for feature space triggers.
Our intuition is that \textit{features representing the Trojan are orthogonal to other features.}
Because a trigger works for a set of samples (or all of them, depending on the attack type), changing the input content without removing the Trojan features will not change the prediction.
That is, changing Trojan and benign features will not affect each other.
Trojan features will form a hyperplane in the high dimensional space, which can constrain the search in feature space.
We then develop our reverse-engineering method by exploiting the feature space constraint.
\autoref{fig:hyperplane_in_feature} demonstrates our idea.
Existing reverse-engineering methods only consider the input space constraint.
It conducts reverse-engineering via searching a static trigger pattern in the input space.
These methods fail to reverse-engineer feature-space Trojans whose trigger is dynamic in the input space.
Instead, our idea is to exploit the feature space constraint and searching a feature space trigger using the constraint that the Trojan features will form a hyperplane.
At the same time, we also reverse-engineer the input space Trojan transformation based on the feature space constraint.
To the best of our knowledge, we are the first to propose feature space reverse-engineering methods for backdoor detection.

Through reverse-engineered Trojans, we developed a Trojan detection and removal method.
We implemented a prototype \sys (\textbf{FEATURE}-space \textbf{RE}verse-engineering) in Python and PyTorch and evaluated it on MNIST, GTSRB, CIFAR, and ImageNet dataset with seven different attacks (i.e., BadNets~\cite{gu2017badnets}, Filter attack~\cite{liu2019abs}, WaNet~\cite{nguyen2021wanet}, Input-aware dynamic attack~\cite{nguyen2020input}, ISSBA~\cite{li2021invisible}, Clean-label attack~\cite{turner2019label}, Label-specific attack~\cite{gu2017badnets}, and SIG attack~\cite{barni2019new}).
Our results show that \sys is effective.
On average, the detection accuracy of our method is 93\%, outperforming existing techniques.
For Trojan mitigation, our method can reduce the ASR (attack success rate) to only 0.26\% with the BA (benign accuracy) remaining nearly unchanged by using only ten clean samples for each class.

Our contributions are summarized as follows.
We first find the feature space properties of the Trojaned model and reveal the relationship between Trojans and the feature space hyperplanes.
We propose a novel Trojan trigger reverse-engineering technique leveraging the feature space Trojan hyperplane.
We evaluate our prototype on four different datasets, five different network architectures, and seven advanced Trojan attacks.
Results show that our method outperforms SOTA approaches.

\section{Background \& Motivation}\label{sec:background}

A DNN classifier is a function \(\mathcal{M} : \mathcal{X}\mapsto \mathcal{Y}\) where \(\mathcal{X}\) is the input domain \(\mathbb{R}^{m}\) and \(\mathcal{Y}\) is a set of labels \(K\).
A Trojan (or backdoor) attack against a DNN model \(\mathcal{M}\) is a malicious way of perturbing the input so that an adversarial input \(x^\prime \) (i.e., input with the perturbation pattern) will be classified to a target/random label while the model maintains high accuracy for benign input \(x\).
The perturbation pattern is known as the Trojan trigger.
Trojan attacks can happen in training (e.g., data poisoning) or model distribution (e.g., changing model weights or supply-chain attack).
Existing works have shown Trojan attacks against different DNN models, including computer vision models~\cite{gu2017badnets,liu2017trojaning,chen2017targeted}, Graph Neural Networks (GNNs)~\cite{xi2021graph, zhang2021backdoor}, Reinforcement Learning (RL)~\cite{kiourti2020trojdrl,wang2021backdoorl}, Natural Language Processing (NLP)~\cite{chen2021badnl,chan2020poison,yang2021careful,qi2021mind,yang2021rethinking,qi2021hidden}, recommendation systems~\cite{zhang2021pipattack}, malware detection~\cite{severi2021explanation}, pretrained models~\cite{shen2021backdoor,yao2019latent,jia2021badencoder}, active learning~\cite{vicarte2021double}, and federated learning~\cite{bagdasaryan2020backdoor,xie2019dba}.
The Trojan trigger can be a simple input pattern (e.g., a yellow pad)~\cite{gu2017badnets,liu2017trojaning,chen2017targeted} or a complex input transformation function (e.g., a CycleGAN to change the input styles)~\cite{cheng2020deep,salem2022dynamic,nguyen2021wanet,nguyen2020input,li2021invisible}.
If the trigger is static input space perturbations (e.g., a yellow pad), the Trojan attack is known as \textit{input-space Trojan}, and if the trigger is an input feature (e.g., an image style), the attack is referred to as the \textit{feature-space Trojan}.

There are different types of Trojan defenses.
A line of work~\cite{chen2018detecting,tran2018spectral,hayase2021defense} attempts to remove poisoned data samples by cleaning the training dataset.
Training-based methods~\cite{li2021anti,wang2022towards,huang2022backdoor} train benign classifiers even with the poisoned dataset.
These training time approaches work for poisoning-based attacks but fail to defend against supply chain attacks where the adversary injects the Trojan after the model is trained.
Another line of work, e.g., STRIP~\cite{gao2019strip}, SentiNet~\cite{chou2018sentinet}, and Februus~\cite{doan2020februus} aim to detect Trojan inputs during runtime.
It is hard to distinguish between a misclassification and a Trojan attack for a test input.
These runtime detection methods make assumptions about the attack, which stronger attacks can violate.
For example, STRIP fails to detect the Trojan inputs when the Trojan trigger locates around the center of an image or overlaps with the main object (e.g., feature space attacks).
Another limitation is that they examine the test inputs and perform various heavyweight tests, significantly delaying the response time.

Trigger reverse engineering~\cite{wang2019neural,liu2019abs,shen2021backdoor,guo2019tabor,chen2019deepinspect,tao2022better,liu2022piccolo,shen2022constrained} makes no assumptions about the attack method (e.g., poisoning or supply-chain attacks) and does not affect the runtime performance.
It inspects the model to check if a Trojan exists before deploying.
Given a DNN model \(\mathcal{M}\) and a small set of clean samples \(\mathcal{X}\), trigger reverse engineering methods try to reconstruct injected triggers.
If reverse engineering is successful, the model is marked as malicious.
Neural Cleanse (NC)~\cite{wang2019neural} proposes to perform reverse engineering by solving \autoref{eq:re}:
\begin{equation}\label{eq:re}
	\mathop{\min}\limits_{\bm{m},\bm{t}} \quad \mathcal{L} \left( \mathcal{M}\left((1-\bm m) \odot \bm{x} + \bm m \odot \bm t \right), y_t\right) + r^{\star}
\end{equation}
where \(x \in \mathcal{X}\) and \(\bm m\) is the trigger mask (i.e., a binary matrix with the same size as the input to determine if the value will be replaced by the trigger or not), \(\bm t\) is the trigger pattern (i.e., a matrix with the same size as the input containing trigger values), and \(r^{\star}\) are attack constraints (e.g., trigger size is smaller than 1/4 of the image). 
\(\mathcal{L}\) is the cross-entropy loss function.
Most prior works~\cite{liu2019abs,shen2021backdoor,guo2019tabor,chen2019deepinspect} follow the same methodology and inherently suffer from the same limitations.
First, they assume that an input space perturbation, denoted by \((\bm{m}, \bm{t}) \), can represent a trigger.
This assumption is valid for input-space triggers but does not hold for feature space attacks.
Second, \(r^{\star}\) are heuristics observed from existing attacks.
For example, NC observed that most triggers have small sizes and limit the trigger size to be no larger than a threshold value.
Otherwise, the trigger will overlap with the main object and decrease benign accuracy.
In practice, more advanced attacks can break such heuristics.
For instance, DFST~\cite{cheng2020deep} leverages CycleGAN to transfer images from one style to another without changing its semantics.
It changes almost all pixels in a given image.
This paper proposes a novel reverse engineering method that overcomes the limitations above for image classifiers.
\section{Methodology}\label{sec:design}

\subsection{Threat Model}\label{sec:threat}
This work aims to determine if a given model has a Trojan or not by reverse-engineering the corresponding trigger.
Following existing works~\cite{wang2019neural,liu2019abs,xu2019detecting}, we assume access to the model and a small dataset containing correctly labeled benign samples of each label.
In practice, such datasets can be gathered from the Internet.
We make no assumptions on how the attacker injects the Trojan (poisoning or supply-chain attack).
The attack can be formally defined as: \(\mathcal{M}(\bm{x}) = y, \mathcal{M}(F(\bm{x})) = y_T, \bm{x} \in \mathcal{X}\), where \(\mathcal{M}\) is the Trojaned model, \(\bm{x}\) is a clean input sample, and \(y_T\) is the target label.
\(F\) is the function to construct Trojan samples.
Input-space triggers add static input perturbations, and feature space triggers are input transformations.
The key difference between our work and existing work is that we consider the feature space triggers.

\subsection{Observation}\label{sec:obs}

In DNNs, the neuron activation values represent its functionality.
The input neurons denote the input space features, and inner neurons extract inner and more abstract features.
Existing reverse-engineering methods constrain the optimization problem in the input space using domain-specific constraints or observations.
For image classification tasks, the pixel value of each image has to be a valid RGB value.
Methods like NC observe that the trigger size must be smaller and cannot overlap with the main object and propose corresponding constraints.
The most challenging problem for reverse-engineering feature space triggers is how to constrain the optimization properly.
Note that there exist a set of neurons; when activating to specific values, the Trojan behavior will be triggered.
Due to the black-box nature of DNNs, it is hard to identify which neurons are related to the Trojan behavior.
Moreover, if enlarge the weight values with the same scale, the output of the DNN will be the same, and as such, it is hard to constrain concrete activation values.
Without a proper constraint, we cannot form an optimization problem.

\begin{figure}[htbp]

    \centering
    \includegraphics[width=1\columnwidth]{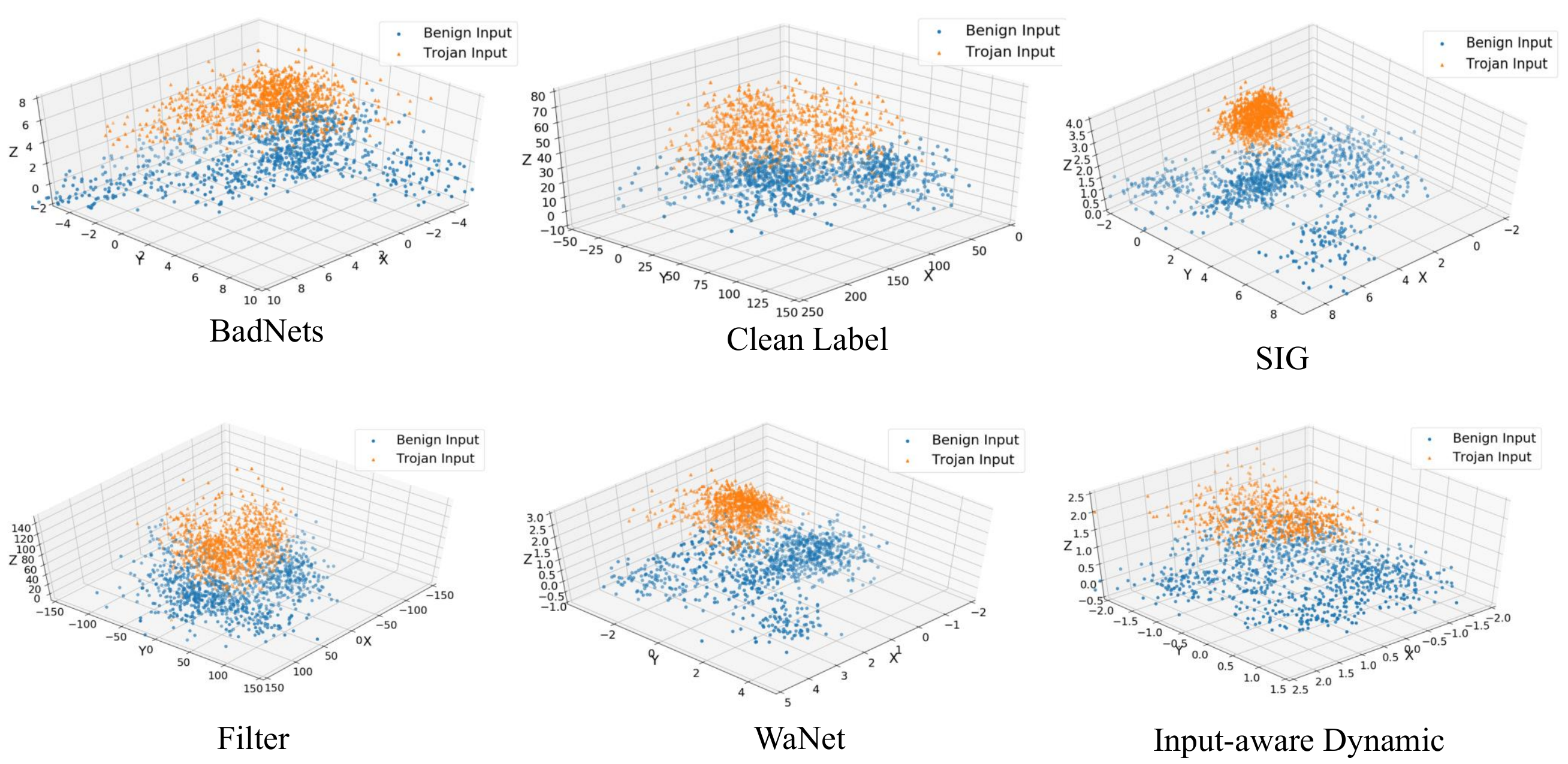}
    \caption{Feature space of Trojaned models.}\label{fig:observation}

\end{figure}

Our key observation to solve this problem is that \textit{neuron activation values representing the Trojan behavior are orthogonal to others}.
Recall that one property of DNN Trojans is that when adding the trigger to \textit{any} given input, the model will predict the output to a specific label.
That is, the trigger will always work regardless of the actual contents of the input.
In the feature space, when the model recognizes features of the Trojan, it will predict the label to the target label regardless of the other features.
These activation values will form a hyperplane space in the high dimensional space so that they can be orthogonal to all others.
Based on this intuition, we performed empirical experiments to confirm our idea.
Specifically, we first use six Trojan attacks (e.g., BadNets~\cite{gu2017badnets}, Clean label attack~\cite{turner2019label}, Filter attack~\cite{liu2019abs}, and WaNet~\cite{nguyen2021wanet}, SIG~\cite{barni2019new} and Input-aware dynamic attack~\cite{nguyen2020input}) to generate Trojaned ResNet18 models on CIFAR-10. We then visualize the feature space of the last convolutional layers in these models. 
In \autoref{fig:observation}, three dimensions, X, Y, and Z, represent the feature space.
We first apply PCA to get two eigenvectors of the benign training set; 
then, we use the obtained eigenvectors as X-axis and Y-axis.
For Z-axis, we first construct Trojan inputs to activate the model's Trojan behavior and find highly related neurons to Trojans.
Then, we use DNN interpretability techniques SHAP~\cite{lundberg2017unified} to estimate the neuron's importance to the Trojan behavior. 
The neurons among the top 3\% are \textit{compromised neurons}.
Z-axis denotes the activation values of compromised neurons.
Namely, 
\(z = \|\mathcal{A}(F(\bm x)) \odot \bm m\|\)
, where $\bm m$ denotes a mask revealing the position of compromised neurons. 
\autoref{fig:observation} show that most Trojan inputs have a similar z-value. They form a linear hyperplane in the feature space while benign ones do not.

\subsection{Feature Space Trojan Hyperplane Reverse-engineering}
In this paper, We use \(\mathcal{A}\) to represent the submodel from the input space to the feature space. \(\mathcal{B}\) is the submodel from the feature space to the output space. 
We also use \(\bm a = \mathcal{A}(\bm x)\) to denote the inner features of the model.
Similar to the reverse-engineering in the input space, given a model \(\mathcal{M}\) and a small set of benign inputs \(\mathcal{X}\), we use a feature space mask \(\bm m\) and a feature space pattern \(\bm t\) to represent the feature space Trojan hyperplane \(H = \{\bm{a}|\bm{m}\odot\bm{a} = \bm{m}\odot\bm{t}\}\).
Specifically, we can update \(\bm m\) and \(\bm t\) via the following optimization process:
\(\mathop{\min}\limits_{\bm{m},\bm{t}} \mathcal{L} \left( \mathcal{B}\left((1-\bm m) \odot \bm a + \bm m \odot \bm t \right), y_t\right)\).
\(y_t\) is the target label.
As discussed above, reverse-engineering the feature space is challenging.
In the input space, all values have natural physical semantics and constraints, e.g., a pixel value in the RGB value range. 
Values in the feature space have uninterruptible meanings and are not strictly constrained.
Whether the result will have a physically meaningful semantic is also uncertain.
We solve these challenges by simultaneously optimizing the input space trigger function \(F\) and the feature space Trojan hyperplane \(H\) to enforce that the trigger has semantic meanings.
In detail, we compute the feature space trigger pattern as the mean inner features on the samples generated by the trigger function, i.e., \(\bm t = mean \left( \bm m \odot \mathcal{A}(F(\mathcal{X}) \right) \).
We also constrain the standard deviation of \(\bm m \odot \mathcal{A}(F(\mathcal{X})\) to make sure the features generated by the trigger function will lie on the relaxation of the reverse-engineered hyperplane.
Formally, our reverse-engineering can be written as the constrained optimization problem shown in \autoref{eq:optimize}, where \(\mathcal{X}\) is the small set of clean samples.
We use deep neural networks to model the trigger function (i.e., \(F = G_{\theta}\)) because of their expressiveness~\cite{chen2019deepinspect,hornik1989multilayer}.
Specifically, we use a representative deep neural network UNet~\cite{ronneberger2015u}. 
Given a model and a small set of clean inputs, the trigger function can be smoothly reconstructed via gradient-based methods, i.e., optimizing the generative model \(G_{\theta}\).
In our default setting, \(\mathcal{A}\) and \(\mathcal{B}\) are separated at the last convolutional layer. 
More discussions are in the Appendix (\autoref{sec:split}).

\begin{equation}
	\begin{split}
		& \mathop{\min}\limits_{F,\bm{m}} \text{ } \mathcal{L} \left(\mathcal{B} \left((1-\bm m) \odot \bm a + \bm m \odot \bm t \right), y_t\right)\\ 
		& \text{where } \bm{t} = \overline{\bm m \odot \mathcal{A}(F(\mathcal{X})} \text{, } \bm a \in \mathcal{A}(\mathcal{X}) \\
		& s.t. \quad \|F(\mathcal{X}) - \mathcal{X}\| \leq \tau_1,\text{ } std(\bm m \odot \mathcal{A}(F(\mathcal{X}))) \leq \tau_2,\text{ } \|\bm m\| \leq \tau_3
	\end{split}
	\label{eq:optimize}
\end{equation}

\begin{algorithm}[tb]
	\caption{Feature-space Backdoor Reverse-engineering}\label{alg:detection1}
	{\bf Input:} %
	\hspace*{0.05in}
	Model: \(\mathcal{M}\)\\ {\bf Output:} \hspace*{0.05in} Trojaned or Not, Trojaned Pairs \(T\) \begin{algorithmic}[1] \Function {Reverse-engineering}{$\mathcal{M}$} \For{{\rm (target class} \(y_t,\) {\rm source class} \(y_s\) {\rm )} {\rm in} \(K\)} \For{\(e \leq E\)} \State \(\bm x = sample(\mathcal{X}_{y_s})\) \State \(cost_1 = \mathcal{L}\left(\mathcal{B} \left((1-\bm m) \odot \bm a + \bm m \odot \bm t \right), y_t\right)\) 

		\If{ \( \|F(\bm x) - \bm x\| \geq \tau_1 \) } \State \(cost_1 = cost_1 + w_1\cdot\|F(\bm x) - \bm x\|\) \EndIf 

		\If{ \( std(\bm m \odot \mathcal{A}(F(\bm x))) \geq \tau_2 \) } \State \(cost_1 = cost_1 + w_2 \cdot std(\bm m \odot \mathcal{A}(F(\bm x)))\) \EndIf 

		\State \(\Delta_{\theta_F} = \frac{\partial cost_1}{\partial \theta_F}\) \State \(\theta_F = \theta_F - lr_1\cdot \Delta_{\theta_F}\) 

		\State \(cost_2 = \mathcal{L}\left(\mathcal{B} \left((1-\bm m) \odot \bm a + \bm m \odot \bm t \right), y_t\right)\) 

		\If{ \( \|\bm m\| \geq \tau_3 \) } \State \(cost_2 = cost_2 + w_3 \cdot \|\bm m\|\) \EndIf 

		\State \(\Delta_{\bm m} = \frac{\partial cost_2}{\partial \bm m}\) \State \(\bm m = \bm m - lr_2 \cdot \Delta_{\bm m}\) 

		\EndFor \If{ \( ASR \left(\mathcal{B} \left((1-\bm m) \odot \bm a + \bm m \odot \bm t \right), y_t\right) > \lambda \) } \State \(\mathcal{M}\) is a Trojaned model, \State \(T.append((y_s, y_t))\) \EndIf \EndFor 

		\EndFunction \end{algorithmic} \end{algorithm}

There are several constraints in the optimization problem: \ding{172} The transformed samples should be similar to the original image due to the properties of Trojan attacks, i.e., \(\|F(\bm x) - \bm x\| \leq \tau_1\).
Typically, the Trojan samples are visually similar to original samples for stealthy purposes. 
In detail, we use MSE (Mean Squared Error) to calculate the distance between \(F(\bm x)\) and \(\bm x\).
\ding{173}
The Trojan features should lie in the relaxation of the reverse-engineered feature space Trojan hyperplane, i.e., \(\mathbb{P}\left(a \in H^{\star}|\bm x \in F(\mathcal{X}) \right)\) should have high values.
To achieve this goal, we constrain the standard deviation of different Trojan samples' activation values on each pixel in the hyperplane.
\ding{174}
Similar to input space trigger reverse-engineering~\cite{liu2019abs}, we set a bound for the size of the feature space trigger mask, i.e., \(\|\bm m\| \leq \tau_3\).
Here \(\tau_1\), \(\tau_2\), and \(\tau_3\) are threshold values.
We discuss their influence in ~\autoref{sec:eval_ablation}.
The detailed reverse-engineering algorithm can be found in~\autoref{alg:detection1}, where \(K\) is a set containing all possible (source class, target class) pairs of the model.
\sys scans all labels to identify the Trojan target labels.
\(w_1\), \(w_2\) and \(w_3\) are the coefficient values used in the optimization.
Following NC~\cite{wang2019neural}, we adjust them dynamically during optimization to make sure the reverse-engineered Trojan satisfies the constraints. 
\(E\) is the maximal epoch number.
In lines 5-11, it optimizes the trigger function \(F\) and then the mask \(\bm m\) of the feature space hyperplane in lines 12-16.
In the end, we determine the reverse-engineering is successful and the label \(y_t\) is a Trojan target label if the attack success rate of the reversed Trojan is above a threshold value \(\lambda\) (80\% in this paper).

\subsection{Trojan Mitigation} 

After we reverse-engineered the Trojans, we can mitigate it by breaking the reverse-engineered feature space Trojan hyperplane. Based on our observation, the neurons in the feature space Trojan hyperplane are highly related to the Trojan behaviors.
Thus, we can mitigate the Trojans by breaking the hyperplane.
Inspired by Zhao et al.~\cite{zhao2021ai}, we can break it by flipping the neurons on it.
Our neuron-flip process can be written as
\autoref{eq:flip}, where \(\bm m\) is the reverse-engineered feature space mask, \(\bm a\) is the inner features.
\(\bm a_i\) is the activation value on the \(i^{th}\) neuron.
\begin{equation}\label{eq:flip}
	Flip(\bm a) = \begin{cases} -\bm a_i,\text{ when \(\bm a_i\) in \(\bm m\)} \\ \bm a_i,\text{ when \(\bm a_i\) not in \(\bm m\)} & \end{cases}
\end{equation}
The mitigated model \(\mathcal{M}^{\prime}(\bm x) = \mathcal{B}\left(Flip(\mathcal{A}(\bm x))\right)\), where \(\mathcal{A}\) and \(\mathcal{B}\) are submodels of the model.

\section{Experiments and Results}
\label{sec:eval}

We first introduce our experiment setup (\autoref{sec:eval_setup}).
We then evaluate the effectiveness of \sys on Trojan detection (\autoref{sec:effectiveness_detection}) and mitigation tasks (\autoref{sec:effectiveness_mitigation}).
We also evaluate the robustness of \sys against different settings and the impacts of configurable parameters in \sys (\autoref{sec:eval_ablation}). 
In \autoref{sec:split}, we discuss how to split the model. 
The adaptive attack can be found in \autoref{sec:adaptive}.

\subsection{Experiment Setup.}
\label{sec:eval_setup}
We implement \sys with python 3.8 and PyTorch.
All experiments are conducted on a Ubuntu 18.04 machine equipped with 64 CPUs and six GeForce RTX 6000 GPUs. 

\begin{wraptable}{r}{0.4\linewidth}
    \centering
    \scriptsize
    \caption{Overview of datasets.}\label{tab:dataset}
    \begin{tabular}{@{}cccc@{}}
    \toprule
    Dataset              & Sample Size & \#Train & Classes \\ \midrule
    MNIST                & 32*32*1    & 60000   & 10       \\ %
    GTSRB                & 32*32*3    & 39209   & 43       \\ %
    CIFAR-10             & 32*32*3    & 50000   & 10        \\ %
    ImageNet        & 224*224*3  & 100000    & 200         \\ \bottomrule
    \end{tabular}
\end{wraptable}

\noindent
\textbf{Datasets and Models.}
We use four publicly available datasets to evaluate \sys, including MNIST~\cite{lecun1998gradient}, GTSRB~\cite{stallkamp2012man}, CIFAR-10~\cite{krizhevsky2009learning} and ImageNet~\cite{russakovsky2015imagenet}.
We summarize our datasets in~\autoref{tab:dataset}.
We show the dataset names, the size of each input sample, the number of samples and the number of classes in each column.
Details of the datasets can be found in Appendix.
For model architectures, we use LeNet5~\cite{lecun1998gradient}, Preact ResNet18 (PRN18)~\cite{he2016identity}, ResNet18~\cite{he2016deep}, a VGG-style network specified in ULP~\cite{kolouri2020universal}, and a model consists of 4 convolutional layers and 2 dense layers used in Xu et al.~\cite{xu2019detecting}. 
These datasets and models are widely used in Trojan-related researches~\cite{gu2017badnets, liu2017trojaning,cheng2020deep,gao2019strip, wang2019neural, liu2019abs,xu2019detecting, liu2018fine}.

\noindent
\textbf{Evaluation Metrics.}
We measure the effectiveness of the Trojan detection task by collecting the detection accuracy (Acc).
Given a set of models consist of benign and Trojaned models, the Acc is the number of correctly classified models over the number of all models.
We also show detailed number of True Positives (TP, i.e., correctly detected Trojaned models), False Positives (FP, i.e., benign models classified as Trojaned models), False Negatives (FN, i.e., Trojaned models classified as benign models) and True Negatives (TN, i.e., correctly classified benign models).
For the Trojan mitigation task, we evaluate the benign accuracy (BA) and attack success rate (ASR)~\cite{veldanda2020nnoculation}.
BA is the number of correctly classified clean inputs over the number of all clean samples.
ASR is defined as the number of Trojan samples that successfully attack models over the number of all Trojan samples.

\noindent
\textbf{Baselines and Attack Settings.} 
We evaluate the performance of \sys on Trojan detection, and compare the results with four reverse-engineering based Trojan detection methods (i.e., ABS~\cite{liu2019abs}, DeepInspect~\cite{chen2019deepinspect}, TABOR~\cite{guo2019tabor}, and K-arm~\cite{shen2021backdoor}) and two classification based methods (i.e., ULP~\cite{kolouri2020universal} and Meta-classifier~\cite{xu2019detecting}).
For Trojan mitigation task, we compare \sys with two advanced mitigation methods (i.e., NAD~\cite{li2021neural} and I-BAU~\cite{zeng2021adversarial}).
We use the default parameter settings described in the original papers of our baseline methods.
To understand the performance of \sys and existing methods against various attack settings, we evaluate them against BadNets~\cite{gu2017badnets}, Filter Trojans~\cite{liu2019abs}, WaNets~\cite{nguyen2021wanet}, IA (Input-dependent dynamic Trojans)~\cite{nguyen2020input}, Clean-label~\cite{turner2019label}, SIG~\cite{barni2019new} and ISSBA (Invisible sample-specific Trojans)~\cite{li2021invisible} attacks.
These attacks are state-of-the-art attack methods and are widely evaluated in Trojan defense papers~\cite{wang2019neural,liu2019abs,zeng2021adversarial,li2021anti}.
If not specified, we use all-to-one (i.e., single-target) setting for all attacks. 
Label-specific setting is discussed in~\autoref{sec:eval_ablation}.%

\subsection{Effectiveness on Trojan Detection}

\label{sec:effectiveness_detection}

To measure the effectiveness on the Trojan detection task, we generate a set of benign and Trojaned models, and then use \sys and existing Trojan detection methods to classify each model.
We collect the Acc, TP, FP, FN and TN results of each method and compare them.
Specifically, we first evaluate the performance of \sys and compare the results with four state-of-the-art reverse-engineering based detection methods. 
We generate 20 Trojaned models as well as 20 benign models on CIFAR-10 dataset for each attack (i.e., BadNets, Filter Trojan, WaNet and Input-aware dynamic Trojan attack).
For MNIST and GTSRB dataset, we train 10 Trojaned and 10 benign LeNet5~\cite{lecun1998gradient} models on each dataset. 
We then compare \sys with two state-of-the-art classification based detection methods. 
Similarly, we generate 10 benign and 10 Trojaned models, and use Trojan detection methods to classify these models.
Notice that, in all Trojan detection tasks, we assume the defender can only access 10 clean samples for each class, which is a common practice.~\cite{wang2019neural,liu2019abs,shen2021backdoor}
The comparison results of reverse-engineering based methods are shown in~\autoref{tab:effectiveness}. 
The results of two classification based methods are demonstrated in~\autoref{tab:effectiveness_classification_ulp} and \autoref{tab:effectiveness_classification_meta}. 
In each table, we show the detailed settings, including dataset names, network architectures, and attack settings.

\begin{table*}[]
    \centering
    \scriptsize
    \caption{Comparison to reverse-engineering methods.}
    \label{tab:effectiveness}
    \setlength\tabcolsep{1.3pt}
    \scalebox{0.97}{
        \begin{tabular}{@{}cccccccccccccccccccccccccccccccccc@{}}
            \toprule
            \multirow{2}{*}{Dataset}  & \multirow{2}{*}{Network}     & \multirow{2}{*}{Attack} &  & \multicolumn{5}{c}{ABS}  &  & \multicolumn{5}{c}{DeepInspect} &  & \multicolumn{5}{c}{TABOR} &  & \multicolumn{5}{c}{K-arm} &  & \multicolumn{5}{c}{FeatureRE} &  \\ \cmidrule(lr){5-9} \cmidrule(lr){11-15} \cmidrule(lr){17-21} \cmidrule(lr){23-27} \cmidrule(lr){29-33}
                                      &                           &                         &  & TP & FP & FN & TN & Acc  &  & TP   & FP   & FN  & TN  & Acc   &  & TP  & FP & FN & TN & Acc  &  & TP  & FP & FN & TN & Acc  &  & TP & FP & FN & TN & Acc  &  \\ \midrule
            MNIST                     & LeNet5                    & WaNet                   &  & 7  & 2  & 3  & 8  & 75\% &  & 4    & 0    & 6   & 10  & 70\%  &  & 3   & 2  & 7  & 8  & 55\% &  & 5   & 0  & 5  & 10 & 75\% &  & 9  & 1  & 1  & 9  & \textbf{90\%} &  \\ \midrule
            GTSRB                     & PRN18                     & WaNet                   &  & 5  & 0  & 5  & 10 & 75\% &  & 5    & 1    & 5   & 9   & 70\%  &  & 2   & 2  & 8  & 8  & 50\% &  & 4   & 0  & 6  & 10 & 70\% &  & 8  & 0  & 2  & 10 & \textbf{90\%} &  \\ \midrule
            \multirow{4}{*}{CIFAR-10} & \multirow{4}{*}{ResNet18} & BadNets                 &  & 18 & 0  & 2  & 20 & 95\% &  & 20   & 2    & 0   & 18  & 95\%  &  & 20  & 3  & 0  & 17 & 93\% &  & 20  & 0  & 0  & 20 & \textbf{100\%} &  & 20 & 1  & 0  & 19 & 98\% &  \\
                                      &                           & Filter                  &  & 13 & 0  & 7  & 20 & 83\% &  & 6    & 2    & 14  & 18  & 60\%  &  & 5   & 3  & 15 & 17 & 55\% &  & 0   & 0  & 20 & 20 & 50\% &  & 18 & 1  & 2  & 19 & \textbf{93\%} &  \\
                                      &                           & WaNet                   &  & 11 & 0  & 9  & 20 & 78\% &  & 11   & 2    & 9   & 18  & 73\%  &  & 3   & 3  & 17 & 17 & 50\% &  & 9   & 0  & 11 & 20 & 73\% &  & 18 & 1  & 2  & 19 & \textbf{93\%} &  \\
                                      &                           & IA                      &  & 3  & 0  & 17 & 20 & 58\% &  & 4    & 2    & 16  & 18  & 55\%  &  & 3   & 3  & 17 & 17 & 50\% &  & 2   & 0  & 18 & 20 & 55\% &  & 19 & 1  & 1  & 19 & \textbf{95\%} &  \\ \bottomrule
            \end{tabular}}
\end{table*}

\begin{table*}[]
    \begin{minipage}{0.5\linewidth}
    \centering
    \scriptsize
    \caption{Comparison to ULP.}
    \setlength\tabcolsep{1.5pt}
    \label{tab:effectiveness_classification_ulp}
    \begin{tabular}{@{}cclcccccccccccc@{}}
        \toprule
        \multirow{2}{*}{Network} & \multirow{2}{*}{Attack} &  & \multicolumn{5}{c}{ULP}  &  & \multicolumn{5}{c}{FeatureRE} &  \\ \cmidrule(lr){4-8} \cmidrule(lr){10-14}
                                 &                         &  & TP & FP & FN & TN & Acc  &  & TP & FP & FN & TN & Acc  &  \\ \midrule
        VGG                      & WaNet                   &  & 1  & 0  & 19 & 20 & 0.53 &  & 17 & 0  & 3  & 20 & \textbf{93\%} &  \\ \bottomrule
        \end{tabular}
    \end{minipage}
    \begin{minipage}{0.5\linewidth}
        \centering
    \scriptsize
    \caption{Comparison to Meta-classifier.}
    \setlength\tabcolsep{1.5pt}
    \label{tab:effectiveness_classification_meta}
    \begin{tabular}{@{}cclcccccccccccc@{}}
        \toprule
        \multirow{2}{*}{Network} & \multirow{2}{*}{Attack} &  & \multicolumn{5}{c}{Meta Classifier} &  & \multicolumn{5}{c}{FeatureRE} &  \\ \cmidrule(lr){4-8} \cmidrule(lr){10-14}
                                 &                         &  & TP    & FP   & FN   & TN   & Acc    &  & TP & FP & FN & TN & Acc  &  \\ \midrule
        4Conv+2FC                & WaNet                   &  & 16    & 4    & 4    & 16   & 0.80   &  & 18 & 0  & 2  & 20 & \textbf{95\%} &  \\ \bottomrule
        \end{tabular}
    \end{minipage}
\end{table*}

\noindent
\textbf{Comparison to Reverse-engineering based methods.}
From the results in~\autoref{tab:effectiveness}, we observe that \sys achieves the best detection results compared with other methods.
The average Acc of \sys is 93\%, which is %
17\%, 23\%, 35\% and 23\%
higher than those of other defense methods.
The results show the benefit of \sys.
When looking into the generalization of Trojan detection methods, we find that \sys can achieve excellent results on both input-space Trojans (i.e., BadNets) and feature-space Trojans (i.e., Filter, WaNet and IA attacks).
However, the performance of existing reverse-engineering methods on feature-space Trojans (i.e., Filter, WaNet and IA attacks) is significantly worse than the performance on static Trojans. 
\sys archives 94\% average Acc but the Acc of TABOR on feature-space Trojans are only 53\%, 50\% and 50\%, respectively. Moreover, \sys has 15.33 TP on average, but existing methods only have 7.87 TP.
\sys can generalize better than existing work because \sys considers both feature and input space constraints.
Existing methods, on the contrary, only consider the input space constraints.
They can not detect feature-space Trojans whose trigger is complex and input-dependent, and directly classify many Trojaned models with feature-space Trojans as benign.

\noindent
\textbf{Comparison to classification based methods.}
When comparing \sys with classification based methods, we notice that \sys has better Acc, more TPs and TNs than classification based methods ULP and Meta-classifier. 
As demonstrated in~\autoref{tab:effectiveness_classification_ulp} and~\autoref{tab:effectiveness_classification_meta}, the Acc of \sys is 93\% and 95\%, which is 40\% and 15\% higher than those of ULP and Meta-classifier.
Overall, the results indicate that \sys is more effective than classification based methods when detecting Trojaned models. 
Different from \sys, which directly inspects models via analyzing its inherent feature space properties, classification based methods highly depend on the external trained dataset.
Therefore, their results are not as precise as \sys.

\begin{table*}[]
  \centering
  \scriptsize
  \caption{Results on Trojan mitigation task (10 clean samples for each class are used).
  }\label{tab:unlearning}
  \setlength\tabcolsep{3pt}
  \begin{tabular}{@{}ccccccccccccccc@{}}
    \toprule
    \multirow{2}{*}{Dataset}  & \multirow{2}{*}{Network}  & \multirow{2}{*}{Attack} &  & \multicolumn{2}{c}{Undefended} &  & \multicolumn{2}{c}{NAD} &  & \multicolumn{2}{c}{I-BAU} &  & \multicolumn{2}{c}{FeatureRE} \\ \cmidrule(lr){5-6} \cmidrule(lr){8-9} \cmidrule(lr){11-12} \cmidrule(l){14-15} 
                              &                           &                         &  & BA             & ASR           &  & BA         & ASR        &  & BA          & ASR         &  & BA          & ASR        \\ \midrule
    MNIST                     & LeNet5                    & WaNet                   &  & 99.22\%        & 94.52\%       &  & 65.87\%    & 48.21\%    &  & 94.87\%     & \textbf{0.22\%}      &  & \textbf{99.20\%}     & 0.63\%     \\ \midrule
    GTSRB                     & PRN18                     & WaNet                   &  & 99.02\%        & 99.70\%       &  & 70.35\%    & 62.76\%    &  & 91.74\%     & 0.86\%      &  & \textbf{98.42\%}     & \textbf{0.00\%}     \\ \midrule
    \multirow{3}{*}{CIFAR-10} & \multirow{3}{*}{ResNet18} & Filter                  &  & 91.30\%        & 98.98\%       &  & 81.66\%    & 28.23\%    &  & 87.45\%     & 18.22\%     &  & \textbf{91.26\%}     & \textbf{0.29\%}     \\
                              &                           & WaNet                   &  & 91.84\%        & 98.17\%       &  & 83.60\%    & 27.52\%    &  & 87.52\%     & 6.84\%      &  & \textbf{91.79\%}     & \textbf{0.04\%}     \\
                              &                           & IA                      &  & 91.62\%        & 92.44\%       &  & 84.03\%    & 34.00\%    &  & 86.88\%     & 10.33\%     &  & \textbf{91.43\%}     & \textbf{0.38\%}     \\ \bottomrule
    \end{tabular}
\end{table*}

\subsection{Effectiveness on Trojan Mitigation}
\label{sec:effectiveness_mitigation}
We evaluate the effectiveness of \sys on Trojan mitigation and compare the results with state-of-the-art methods NAD and I-BAU. 
We use the Trojaned models generated by three attacks (i.e., Filter attack, WaNet and IA)
and report their average BA and ASR after Trojan mitigation.
We also show the average BA and ASR of undefended Trojaned models. 
For all methods, 
the defenders can access 10 clean samples for each class to conduct Trojan mitigation.
We show the results in~\autoref{tab:unlearning}.

We find that \sys is the most effective method for Trojan mitigation among all methods.
Compared to state-of-the-art Trojan mitigation methods, \sys archives the lowest average ASR and the highest average BA.
On the one hand, using \sys can decrease the average ASR from 96.76\% to 0.26\%. 
Other methods can only decrease the average ASR to 40.14\% and 7.29\%.
The results show the advantages of \sys on Trojan mitigation.
On the other hand, the BA with \sys is similar to undefended models.
But the BA of other methods is significantly lower than that of undefended models.
By breaking the feature space hyperplane, \sys can successfully mitigate Trojans with minimal BA loss.
Other methods, which cannot effectively find Trojan-related features, cannot achieve good results.

\subsection{Ablation Study}
\label{sec:eval_ablation}

In this section, we evaluate the resistance of \sys to various Trojan attack settings and large datasets.
We also evaluate the impacts of configurable parameters in \sys, including the constrain values used in \autoref{eq:optimize} and the number of used clean samples. 
By default, the attack used for measuring the impacts of configurable parameters is IA.
We use 20 benign ResNet models and 20 Trojaned ResNet models on CIFAR-10 to test the detection results.
Notice that we only evaluate the performance on the Trojan detection task.
Due to the page limits, we include the ablation study on Trojan mitigation in Appendix (\autoref{sec:ablation_mitigation}).

\begin{wraptable}{r}{0.45\linewidth}
    \centering
    \scriptsize
    \setlength\tabcolsep{1.5pt}
    \caption{Resistance to more attacks.}
    \scalebox{0.95}{
        \begin{tabular}{@{}cccccccc@{}}
            \toprule
            Dataset                   & Network                      & Attack & TP & FP & FN & TN & Acc  \\ \midrule
            \multirow{3}{*}{CIFAR-10} & \multirow{3}{*}{ResNet18} & LS     & 9  & 1  & 1  & 9  & 90\% \\
                                      &                           & CL     & 8  & 1  & 2  & 9  & 85\% \\
                                      &                           & SIG    & 10 & 1  & 0  & 9  & 95\% \\ \midrule
            ImageNet                  & ResNet18                  & ISSBA  & 4  & 0  & 1  & 5  & 90\% \\ \bottomrule
            \end{tabular}}
    \label{tab:more_settings}
\end{wraptable}
\noindent
\textbf{Resistance to various attack and dataset settings.}
To evaluate if our method is resistant to more Trojan attacks, we train 20 Trojaned ResNet18 models on CIFAR-10 for Label-specific attack (LS), Clean-label attack (CL) and SIG attack (SIG). 
For the label-specific attack, we consider the all-to-all attack setting, i.e., the target label \(y_T = \eta(y) = y+1\), where \(\eta\) is a mapping and \(y\) is the correct label of the sample. 
In addition, we generate five benign models and five Trojaned models with ISSBA attacks on ImageNet to evaluate if our method is compatible with large-scale datasets.
We summarize the results in~\autoref{tab:more_settings}.

In \autoref{tab:more_settings}, we find that \sys is compatible with evaluated Trojan attacks, showing the generalization of our reverse-engineering based method. We also observe that our method has high Acc on the ImageNet dataset with ISSBA~\cite{li2021invisible}. Thus, our method is also applicable to large datasets.

\noindent
\textbf{Influence of constrain values.}
As shown in \autoref{eq:optimize}, there are three constrain values (\(\tau_1\), \(\tau_2\), \(\tau_3\)) in our constrained optimization process.
By default, \(\tau_1 = 0.15\), \(\tau_2 = 0.25\) and \(\tau_3 = 5\%\).
We evaluate their influences.
For \(\tau_1\), we calculate input space perturbations on the preprocessed inputs, and the details of the preprocessing can be found in Appendix (\autoref{sec:appendix_details_datasets}).
We vary \(\tau_1\) from 0.05 to 0.35, change \(\tau_2\) from 0.10 to 0.50, and tune \(\tau_3\) from 3\% of the whole feature space to 10\% of the whole feature space.
The results under different hyperparameter settings are shown in
\autoref{tab:ablation}.

From the results, we observe that the performance of \sys is insensitive to these three hyperparameters. 
In detail, when we vary \(\tau_1\), \(\tau_2\) and \(\tau_3\), the Acc is stable.
In all cases, our method always achieves over 90\% detection accuracy.
The results further show the robustness of \sys.
We also find that, when the value of all hyperparameters becomes lower, \sys has more FN. 
On the contary, when its value is larger, more FP will be produced. 
This is understandable because lower constrain values mean a stricter criterion for a successful reverse-engineering.

\begin{minipage}[t]{\textwidth}
    \begin{minipage}{0.55\textwidth}
        \centering
  \scriptsize
  \setlength\tabcolsep{2pt}
  \captionof{table}{Influence of hyperparameters.}
  \label{tab:ablation}
  \scalebox{1}{
    \begin{tabular}{@{}ccccccccccccc@{}}
        \toprule
        \multirow{2}{*}{Metric} &  & \multicolumn{3}{c}{\(\tau_1\)} &  & \multicolumn{3}{c}{\(\tau_2\)} &  & \multicolumn{3}{c}{\(\tau_3\)} \\ \cmidrule(lr){3-5} \cmidrule(lr){7-9} \cmidrule(l){11-13} 
                                &  & 0.05         & 0.15         & 0.35         &  & 0.10         & 0.25         & 0.50         &  & 3\%          & 5\%          & 10\%         \\ \midrule
        TP                      &  & 18           & 19           & 20           &  & 17           & 19           & 19           &  & 17           & 19           & 19           \\
        FP                      &  & 0            & 1            & 3            &  & 1            & 1            & 2            &  & 0            & 1            & 2            \\
        FN                      &  & 2            & 1            & 0            &  & 3            & 1            & 1            &  & 3            & 1            & 1            \\
        TN                      &  & 20           & 19           & 17           &  & 19           & 19           & 18           &  & 20           & 19           & 18           \\
        Acc                     &  & 0.95         & 0.95         & 0.93         &  & 0.90         & 0.95         & 0.93         &  & 0.93         & 0.95         & 0.93         \\ \bottomrule
        \end{tabular}}
    \end{minipage}
    \hfill
    \begin{minipage}{0.43\textwidth}
        \centering
        \includegraphics[width=0.9\columnwidth]{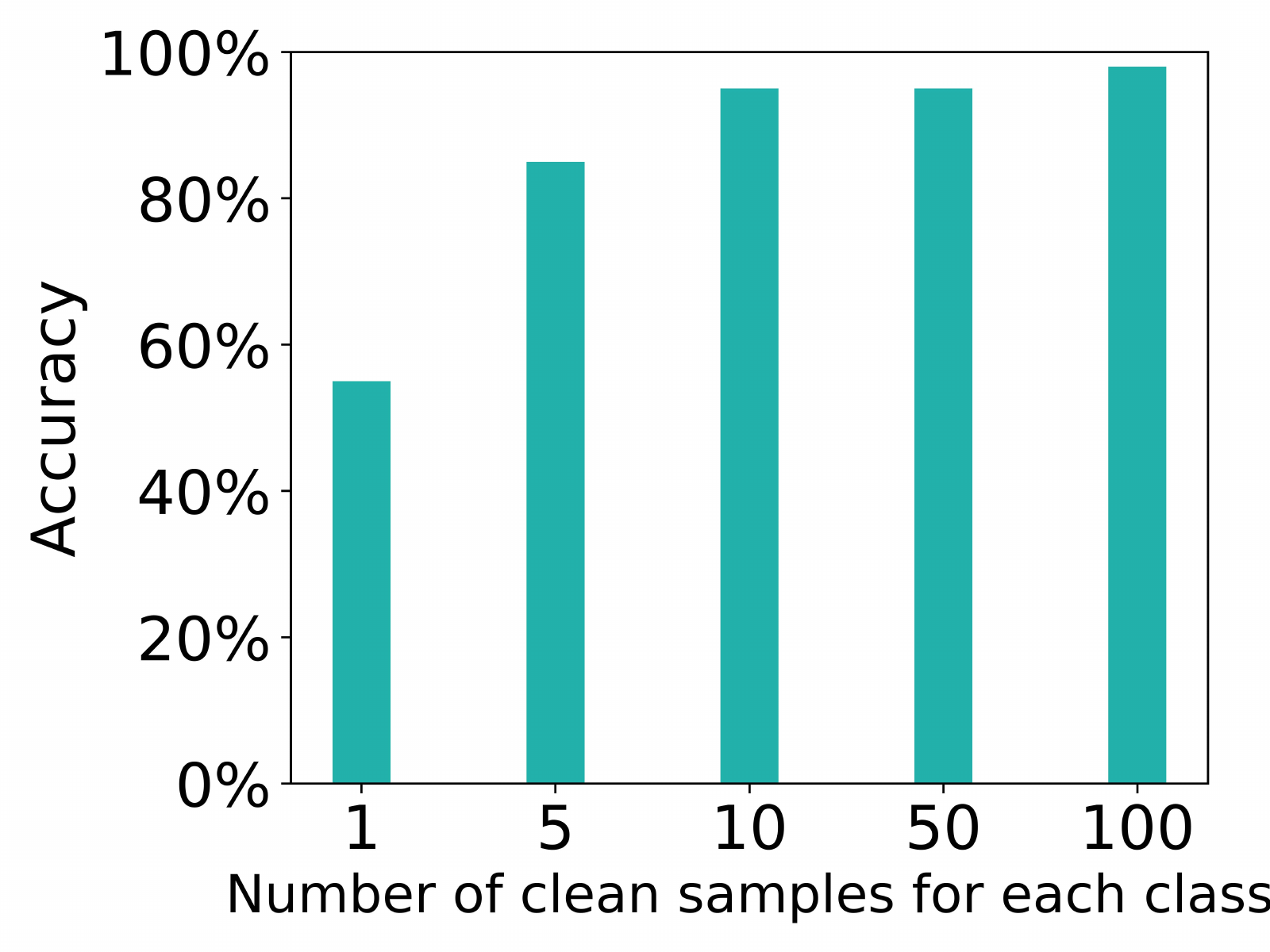}
        \captionof{figure}{Effects of clean set size.
        }
        \label{fig:diff_datanum}
      \end{minipage}
\end{minipage}

\noindent
\textbf{Number of clean reference samples.}
Our threat model and existing work assume the defender can access a set of clean samples for defense.
To investigate the influences of the number of used clean samples in Trojan detection, we choose the number from 1 to 100 in each class and report the Acc results.
The results are shown in~\autoref{fig:diff_datanum}.

From the results, we notice that the Acc decreases significantly when we use less than 10 samples for each class.
This is because the number of used sample affects the optimization process. 
When the number of used samples is too small, the optimization process might be problematic, e.g., it encounters overfitting problem. 
When the number of used samples is larger than 10, \sys achieves high detection accuracy (i.e., above 95\%) and the Acc will not change significantly when the number of used samples keeps increasing.
The reason is using more data makes the optimizing process converge and finally arrives a stable state.
Note that requiring hundreds clean samples is common for reverse-engineering based methods~\cite{wang2019neural,liu2019abs,guo2019tabor} and other types of defenses~\cite{gao2019strip,li2021neural,zeng2021adversarial,li2021anti}.
\sys only requires 10 clean samples for each class, which is more efficient.

\subsection{Discussion for Model Split}\label{sec:split}
As we discussed in \autoref{sec:design}, our method split the model \(\mathcal{M}\) into two sub-models \(\mathcal{A}\) and \(\mathcal{B}\).
In this section, we discuss the influence of using different split positions.
\autoref{tab:split} shows the results of using different \(\mathcal{A}\) and \(\mathcal{B}\) on the ResNet18 model and CIFAR-10 dataset.
In detail, we report the results of splitting the 
model at the \nth{9}, \nth{11}, \nth{13}, \nth{15}, and the last convolutional layer.
The average detection accuracy for splitting at the \nth{9} layer, \nth{11} layer, \nth{13} layer, \nth{15} layer, and last layer is 86.50\%, 87.75\%, 89.50\%, 94.00\%, and 94.75\%, respectively.
As we can see, the performance of splitting at later layers is higher than the performance of splitting at earlier layers.
In our current implementation, we set \(A(x)\) as the sub-model from the input layer to the last convolution layer and \(B(x)\) as the rest.
The relationship between the input and the output of a convolutional layer \(L_n\) is \(x_{n+1} = L_n(x_n) = \sigma (\mathbf{W}_{n}^\mathbf{T}x_n+\mathbf{b}^\mathbf{T}_n)\), where \(x_n\) and \(x_{n+1}\) are the inputs and outputs of layer \(n\), \(\mathbf{W}_n\) and \(\mathbf{b}_n\) are weights and bias values, and \(\sigma \) is the activation function.

\begin{minipage}[htbp]{\textwidth}
    \begin{minipage}{0.5\textwidth}
        \centering
          \scriptsize
          \captionof{table}{Accuracy on different split position.}
          \label{tab:split}
          \scalebox{1}{
            \begin{tabular}{@{}cccccc@{}}
            \toprule
            Attack  & \nth{9}  & \nth{11} & \nth{13} & \nth{15} & Last \\ \midrule
            BadNets & 88\% & 93\% & 98\% & 98\% & 98\% \\
            Filter  & 88\% & 85\% & 90\% & 95\% & 93\% \\
            WaNet   & 85\% & 88\% & 85\% & 93\% & 93\% \\
            IA      & 85\% & 85\% & 85\% & 90\% & 95\% \\ \bottomrule
            \end{tabular}}
            \end{minipage}
    \begin{minipage}{0.5\textwidth}
        \centering
          \scriptsize
          \captionof{table}{Results on BadNets and adaptive attack.}
          \label{tab:adaptive}
          \scalebox{1}{
        \begin{tabular}{@{}cccc@{}}
    \toprule
    Attack   & BA      & ASR     & Acc \\ \midrule
    BadNets  & 94.34\% & 99.98\% & 98\%               \\
    Adaptive & 87.36\% & 93.67\% & 65\%               \\ \bottomrule
    \end{tabular}}
        \end{minipage}
\vspace{0.4cm}
\end{minipage}

Based on existing literatures~\cite{zhou2018interpretable,bau2020understanding}, the features in the deeper CNN layers are more disentangled than that of earlier layers.
Thus, if the orthogonal phenomenon happens in a layer \(L_{n}\), it will exist for all its subsequent layers, e.g., \(L_{n+1}\).
If the orthogonal phenomenon does not happen, the layer without this phenomenon will mix benign and backdoor features, leading to low benign accuracy or attack success rate.
The results in \autoref{tab:adaptive} confirm our analysis.
Thus, a successful backdoor attack will lead to the orthogonal phenomenon in the last convolution layer.

\vspace{-0.2cm}

\subsection{Adaptive Attack}\label{sec:adaptive}
\vspace{-0.2cm}

Our threat model assumes that the attacker can control the training process of the Trojan model.
In this section, we discuss the potential adaptive attacker that knows our defense strategy and tries to bypass \sys via modifying the training process.
Our observation is that the neuron activation values representing the Trojan behavior are orthogonal to others.
One possible adaptive attack is breaking such orthogonal relationships during the Trojan injection process.
We design an adaptive attack that adds one loss term to push the Trojan features to be not orthogonal to benign features.
This attack can be formulated as: \(L = L_{ce} + L_{adv}\), where \(L_{ce}\) is the standard classification loss and the \(L_{adv}\) is defined as:

\vspace{-0.2cm}
\begin{equation}\label{eq:adaptive}
	L_{adv} = \text{SIM}(\mathcal{B}(m\odot a + (1-m) \odot t), \mathcal{B}(m\odot a^{\prime} + (1-m) \odot t)) 
\end{equation} 
\vspace{-0.2cm}

Here, \(\text{SIM}\) is the cosine similarity;
\(a\) and \(a^{\prime}\) are the features of different benign samples; 
\(m\) and \(t\) are the feature-space mask and pattern of the compromised neurons obtained via SHAP~\cite{lundberg2017unified}.
The loss term \(L_{adv}\) tries to enforce the Trojan features being not orthogonal to the benign ones.
We conduct this adaptive attack on the CIFAR-10 dataset and ResNet18 model.
The results can be found in \autoref{tab:adaptive}.
The detection accuracy of \sys under adaptive attack drops to 65\%.
Meanwhile, the average BA/ASR of the adaptive attack and BadNets (native training) is 87.36\%/94.34\% and 93.67\%/99.98\%, respectively.
The adaptive attack can reduce the detection accuracy of our method.
Both the BA and ASR of the adaptive attack are lower than those of native training.
The results confirm our analysis in \autoref{sec:split}: the model without the ``orthogonal phenomenon'' will mix benign and Trojan features, leading to low benign accuracy or attack success rate.
\vspace{-0.2cm}

\section{Discussion}\label{sec:discussion}
\vspace{-0.2cm}

\textbf{Limitations of our method.}
Similar to most existing Trojaned model detection and mitigation methods~\cite{wang2019neural,liu2019abs,liu2022piccolo,shen2021backdoor,shen2022constrained,guo2019tabor,chen2019deepinspect,liu2018fine,li2021neural,tao2022better}, our method requires a small set of clean samples.
In the real world, these samples can be obtained from the Internet.

\textbf{Ethics.}
This paper proposes a technique to detect and remove Trojans in DNN models.
We believe it will help improve the security of DNNs and be beneficial to society.

\section{Conclusion}
\label{sec:conclusion}
\vspace{-0.2cm}

In this paper, we
find relationships between feature space hyperplane and Trojans in DNNs. 
More over, we propose a new Trojaned DNN detection and mitigation method based on our findings. Compared to the state-of-the-art methods, our
method has better performance in both detection and mitigation tasks.

\section*{Acknowledgement}
\vspace{-0.2cm}

We thank the anonymous reviewers for their constructive comments.
This work is supported by IARPA TrojAI W911NF-19-S-0012.
Any opinions, findings, and conclusions expressed in this paper are those of the authors only and do not necessarily reflect the views of any funding agencies.

\bibliographystyle{unsrtnat}
\bibliography{reference}

\section*{Checklist}

\begin{enumerate}

\item For all authors...
\begin{enumerate}
  \item Do the main claims made in the abstract and introduction accurately reflect the paper's contributions and scope?
    \answerYes{}
  \item Did you describe the limitations of your work?
    \answerYes{See \autoref{sec:discussion}.}
  \item Did you discuss any potential negative societal impacts of your work?
    \answerYes{See \autoref{sec:discussion}.}
  \item Have you read the ethics review guidelines and ensured that your paper conforms to them?
    \answerYes{See \autoref{sec:discussion}.}
\end{enumerate}

\item If you are including theoretical results...
\begin{enumerate}
  \item Did you state the full set of assumptions of all theoretical results?
    \answerNA{}
	\item Did you include complete proofs of all theoretical results?
    \answerNA{}
\end{enumerate}

\item If you ran experiments...
\begin{enumerate}
  \item Did you include the code, data, and instructions needed to reproduce the main experimental results (either in the supplemental material or as a URL)?
    \answerYes{See Abstract, \autoref{sec:eval}, and Appendix.}
  \item Did you specify all the training details (e.g., data splits, hyperparameters, how they were chosen)?
    \answerYes{See \autoref{sec:eval} and Appendix.}
	\item Did you report error bars (e.g., with respect to the random seed after running experiments multiple times)?
    \answerNo{}
	\item Did you include the total amount of compute and the type of resources used (e.g., type of GPUs, internal cluster, or cloud provider)?
    \answerYes{See \autoref{sec:eval}.}
\end{enumerate}

\item If you are using existing assets (e.g., code, data, models) or curating/releasing new assets...
\begin{enumerate}
  \item If your work uses existing assets, did you cite the creators?
    \answerYes{See \autoref{sec:eval}.}
  \item Did you mention the license of the assets?
    \answerYes{See Appendix.}
  \item Did you include any new assets either in the supplemental material or as a URL?
    \answerYes{See Abstract.}
  \item Did you discuss whether and how consent was obtained from people whose data you're using/curating?
    \answerYes{See Appendix.}
  \item Did you discuss whether the data you are using/curating contains personally identifiable information or offensive content?
    \answerYes{See Appendix.}
\end{enumerate}

\item If you used crowdsourcing or conducted research with human subjects...
\begin{enumerate}
  \item Did you include the full text of instructions given to participants and screenshots, if applicable?
    \answerNA{}
  \item Did you describe any potential participant risks, with links to Institutional Review Board (IRB) approvals, if applicable?
    \answerNA{}
  \item Did you include the estimated hourly wage paid to participants and the total amount spent on participant compensation?
    \answerNA{}
\end{enumerate}

\end{enumerate}

\newpage
\appendix
\section{Appendix}\label{sec:appendix}

\noindent
{\bf Roadmap:}
More details of Algorithm 1 is introduced in \autoref{sec:more_details_alg}.
Then, we present more details of the datasets (\autoref{sec:appendix_details_datasets}) and attacks (\autoref{sec:appendix_details_attacks}) used in the experiments.
We also perform an ablation study for the Trojan mitigation task in \autoref{sec:ablation_mitigation}.
In \autoref{sec:eval_visualization}, we visualize our reverse-engineered Trojans.
We also show the generalization (\autoref{sec:generalization}) and the efficiency (\autoref{sec:efficiency}) of \sys.
Finally, we discuss the findings in the evaluation (\autoref{sec:findings_evaluation}).

\subsection{More details of Algorithm 1}\label{sec:more_details_alg}
In this section, we discuss more details of our Reverse-engineering Algorithm (\autoref{alg:detection1}).
Given a model \(\mathcal{M}\) and a small set of clean samples \(\mathcal{X}\), the output of the algorithm is a flag indicating if the model is Trojaned, and Trojaned label pairs denoting the source label and the target label of the detected Trojans.

In line 2, we iterate (source label, target label) pair from possible pairs \(K\).
\(E\) in line 3 means the maximal optimization epoch number for each pair.
It is set to 400 in this paper.
In line 4, we randomly sample a batch of inputs from the samples in source classes.
The batch size is set to 128 by default.

In lines 5 to Line 11, we optimize the parameters of the input space transformation \(F\), which is represented by a UNet~\cite{ronneberger2015u} model in our implementation.
In line 5, we calculate the loss value specified in \autoref{eq:re}, where \(\bm a = \mathcal{A}(\bm x)\) is the inner feature on clean samples.
By default, \(\mathcal{A}\) is the submodel from the input layer to the penultimate layer, and \(\mathcal{B}\) is the submodel from the penultimate layer to the output layer.
\(\bm m\) is the feature space trigger mask.
\(\bm{t} = mean \left(\bm m \odot \mathcal{A}(F(\mathcal{X})\right)\) is the feature space trigger pattern. \(\mathcal{L}\) is the cross-entropy loss calculating the distance between the target label and the output of the model under inner features with feature space Trojans.
In line 6, if the input space MSE (Mean Square Error) distance for original inputs \(\bm x\) and the transformed inputs \(F(\bm x)\) is larger than a threshold value \(\tau_1\) (i.e., 0.15), then the regularization item \(w_1\cdot\|F(\bm x) - \bm x\|\) will be added.
Note that we calculate input space distance on the preprocessed inputs, and the details of the preprocessing are in \autoref{sec:appendix_details_datasets}.
Following NC~\cite{wang2019neural}, the coefficient value \(w_1\) is adjusted dynamically to make the reverse-engineering satisfy the constrain (i.e., \(\|F(\bm x) - \bm x\| \leq \tau_1\)).
\(w_2\) in line 9 and \(w_3\) in line 14 are also adjusted dynamically.
In lines 8-9, similarly, we add the regularization item for the standard deviation of different Trojan samples' activation values on each pixel in the hyperplane.
The default value for \(\tau_2\) is 0.25.
Lines 10-11 are the standard backward propagation process to update the parameters of the input space transformation function \(F\) based on the gradients.
The optimizer used to optimize \(F\) is Adam~\cite{kingma2014adam}.
The value of learning rate \(lr_1\) is 1e-3.
In each epoch, we optimize both the input space transformation function \(F\) and the feature space mask \(\bm m\).

Lines 12-16 describes the process for optimizing \(\bm m\).
Similar to line 5, we calculate the cross-entropy loss between the target label and the output of the model under inner features with feature space Trojans in line 12.
In lines 13-14, we add the regularization item for the size of the feature space Trojan hyperplane.
The default value for \(\tau_3\) is 5\% of the whole feature space.
Lines 15-16 describe the process of updating feature space mask \(\bm m\) via gradients.
The value of learning rate \(lr_2\) in line 16 is 1e-1.
The optimizer used is Adam~\cite{kingma2014adam}.

In line 17, we check if the Trojan is successfully reverse-engineered.
In detail, we calculated the ASR (attack success rate) on inner features with feature space Trojans (i.e., \((1-\bm m) \odot \bm a + \bm m \odot \bm t\)).
We flag that reverse-engineering is successful if the ASR is above a threshold value \(\lambda\) (i.e., 0.8).
If the Trojan is successfully reverse-engineered, we flag the model as a Trojan model and label the (source class, target class) pair as Trojaned pair.
Besides the details above, we also use K-arm scheduler~\cite{shen2021backdoor} to speed up the reverse engineering.
Lastly, we use Liu et al.~\cite{liu2022complex} to distinguish the Injected Trojans and UAPs (Universal Adversarial Patterns)~\cite{moosavi2017universal}.

\subsection{Details of Datasets}\label{sec:appendix_details_datasets}

In this section, details of the used
 datasets are discussed.
We also provide the details of the preprocessing for each dataset. All datasets are open-sourced.
The license for all datasets is the MIT license.
They do not contain any personally identifiable information or offensive content.

\smallskip
\noindent
\textbf{MNIST~\cite{lecun1998gradient}.}
This dataset is used for classifying hand-written digits.
It contains 60000 training samples in 10 classes.
The number of samples in the test set is 10000.

\smallskip
\noindent
\textbf{GTSRB~\cite{stallkamp2012man}}
This dataset is built for traffic sign classification tasks.
The number of classes is 43.
The sample numbers for the training set and test set are 39209 and 12630, respectively.

\smallskip
\noindent
\textbf{CIFAR10~\cite{krizhevsky2009learning}}
This dataset is used for recognizing general objects, e.g., dogs, cats, and planes.
It has 50000 training samples and 10000 training samples.
This dataset has 10 classes.

\smallskip
\noindent
\textbf{ImageNet~\cite{russakovsky2015imagenet}}
This dataset is also a general object classification benchmark.
Note that we use a subset (containing 200 classes) of the original ImageNet dataset specified in ISSBA~\cite{li2021invisible}.
The subset has 100000 training samples and 10000 test samples.

Following standard convention on the image classification task, we scale the inputs to the range [0,1] and use mean-std normalization to preprocess the images.
In detail, the preprocessing can be written as \(\bm x^{\prime} = \frac{(\frac{\bm x}{255} - Mean)}{Std}\), where \(\bm x^{\prime}\) is the normalized input and \(\bm x\) is the original inputs.
The Mean value and Std (Standard Deviation) value for each channel on different datasets are summarized in \autoref{tab:preprocess}.

\begin{table}[]
    \centering
    \scriptsize
    \caption{Details of Mean and Std value on each dataset.}\label{tab:preprocess}
    \vspace{0.1cm}
    \begin{tabular}{@{}ccc@{}}
    \toprule
    Dataset  & Mean                         & Std                          \\ \midrule
    MNIST    & {[}0.1307{]}                 & {[}0.3081{]}                 \\
    CIFAR-10 & {[}0.4914, 0.4822, 0.4465{]} & {[}0.2023, 0.1994, 0.2010{]} \\
    GTSRB    & {[}0.3403, 0.3121, 0.3214{]} & {[}0.2724, 0.2608, 0.2669{]} \\
    ImageNet & {[}0.4850, 0.4560, 0.4060{]}    & {[}0.2290, 0.2240, 0.2250{]}    \\ \bottomrule
    \end{tabular}
\end{table}

\subsection{Details of Attacks}\label{sec:appendix_details_attacks}

In this section, we discuss the details of the used attacks.
By default, the attacks are in all-to-one (i.e., single-target) setting, and the target label is randomly selected when we generate Trojaned models.

\noindent
\textbf{BadNets~\cite{gu2017badnets}.}
This attack uses a fixed pattern (i.e., a patch or a watermark) as Trojan triggers, and it generates Trojan inputs by simply pasting the pre-defined trigger pattern on the input.
It compromised the victim models by poisoning the training data (i.e., injecting Trojan samples and modifying their labels to target labels).
In our experiments, we use a 3*3 yellow patch located at the left-upper corner as Trojan trigger.
The poisoning rate we used is 5\%.
The attack can be all-to-one (i.e., single-target) and all-to-all (i.e., label-specific).
For an all-to-one attack, all Trojan samples have the same target label.
For label-specific attacks, the samples in different original classes have different target labels.
In our experiment, the target label for label-specific attack is \(y_T = \eta(y) = y+1\), where \(\eta\) is a mapping and \(y\) is the correct label of the sample.

\noindent
\textbf{Filter Attack~\cite{liu2019abs}.}
This attack exploits image filters as triggers and creates Trojan samples by applying selected filters on images.
Similar to BadNets, the Trojans are injected with poisoning.
Following ABS~\cite{liu2019abs}, we use a 5\% poisoning rate and apply the Nashville filter from Instagram as the Trojan trigger.

\noindent
\textbf{WaNet~\cite{nguyen2021wanet}.}
This method achieves Trojan attacks via image warping techniques.
The trigger transformation of this attack is an elastic warping operation.
Different from BadNets and Filter Attack, in this attack, the adversary needs to modify the training process of the victim models to make the attack more resistant to Trojan defenses.
It is stealthy to human inspection, and it can also bypass many existing Trojan defense mechanisms~\cite{chen2018detecting,gao2019strip,wang2019neural,liu2018fine}.
In our experiments, the wrapping strength and the grid size are set to 0.5 and 4, respectively.

\noindent
\textbf{Input-aware Dynamic Attack~\cite{nguyen2020input}.}
This attack generates Trojan triggers via a trained generator network.
The trigger generator is trained on a diversity loss so that two different input images do not share the same trigger.
Similar to WaNet~\cite{nguyen2021wanet}, the attacker needs to control the training process.

\noindent
\textbf{SIG~\cite{barni2019new}.}
This method uses superimposed sinusoidal signals as Trojan triggers.
In this attack, the attacker can only poison a set of training samples but can not control the full training process.
We set the poisoning rate as 5\%.
The frequency and the magnitude of the backdoor signal in our experiments are 6 and 20, respectively.

\noindent
\textbf{Clean Label Attack~\cite{turner2019label}.}
This attack poisons the datasets without manipulating the label of poisoning samples so that the attack is more stealthy.
The poisoning samples are generated by a trained GAN.
In our experiments, we set the poisoning rate as 5\%.

\noindent
\textbf{ISSBA~\cite{li2021invisible}.}
This attack utilizes an encoder-decoder network to generate sample-specific triggers.
The generated triggers are invisible noises.
The generated noises also contain the information of a representative string of the target label.
The threat model of this attack is that the attacker can only poison the training data, but can not control other components in training (e.g., the loss function).
Following the original paper, we poison 10\% training data in our experiments.

\subsection{Ablation Study on Trojan Mitigation}
\label{sec:ablation_mitigation}

In this section, we study the performance of \sys under different constrain values and different numbers of used clean samples.
The attack used in this section is WaNet~\cite{nguyen2021wanet}.

\noindent
\textbf{Influence of constrain values.}
To investigate the influence constrain values (i.e., \(\tau_1\), \(\tau_2\), and \(\tau_3\)) on the Trojan mitigation performance, We vary \(\tau_1\) from 0.05 to 0.35, change \(\tau_2\) from 0.10 to 0.50, and tune \(\tau_3\) from 1\% of the whole feature space to 7\% of the whole feature space.
We collect the BA and ASR of the mitigated models and report them in \autoref{tab:ablation_mitigation}.
The results show that the mitigation performance of \sys is not sensitive to \(\tau_1\) and \(\tau_2\).
For \(\tau_3\), when the size of the Trojan hyperplane is extremely small (e.g., 1\% of the feature space), the ASR is high.
This is understandable because breaking an extremely small feature space Trojan hyperplane means flipping a very small number of neurons, and it is not enough to completely remove the Trojans in the model.
Therefore, we set the default value of the hyperplane's size as 5\% of the feature space.

\begin{table}[]
    \centering
    \scriptsize
    \setlength\tabcolsep{2pt}
    \caption{Influence of hyperparameters on Trojan mitigation task.}
    \vspace{0.1cm}
    \label{tab:ablation_mitigation}
    \begin{tabular}{@{}clccclccclcccccccl@{}}
\toprule
\multirow{2}{*}{Metric} &  & \multicolumn{3}{c}{$\tau_1$}  & \multicolumn{1}{c}{} & \multicolumn{3}{c}{$\tau_2$}  & \multicolumn{1}{c}{} & \multicolumn{7}{c}{$\tau_3$}                                          & \multicolumn{1}{c}{} \\ \cmidrule(lr){3-5} \cmidrule(lr){7-9} \cmidrule(lr){11-17}
                        &  & 0.05    & 0.15    & 0.35    & \multicolumn{1}{c}{} & 0.10    & 0.25    & 0.50    &                      & 1\%    & 2\%    & 3\%    & 4\%    & 5\%    & 6\%    & 7\%    &                      \\ \cmidrule(r){1-17}
BA                      &  & 91.77\% & 91.79\% & 91.79\% &                      & 91.76\% & 91.79\% & 91.80\% &                      & 91.92\% & 91.87\% & 91.85\% & 91.82\% & 91.79\% & 91.65\% & 90.08\% &                      \\
ASR                     &  & 0.02\%  & 0.04\%  & 0.08\%  &                      & 0.02\%  & 0.04\%  & 0.08\%  &                      & 57.75\% & 0.50\%  & 0.06\%  & 0.06\%  & 0.04\%  & 0.00\%  & 0.00\%  &                      \\ \bottomrule
\end{tabular}
\end{table}

\noindent
\textbf{Number of clean reference samples.}
To understand the influence of clean set size on the Trojan mitigation task, we vary the number of used clean samples from 5 per class to 100 per class and report the BA and ASR of mitigated model.
The results in \autoref{tab:ablation_mitigation_datanum} demonstrate that the performance of \sys is robust when the number of used samples changes.

\begin{table}[]
    \centering
    \scriptsize
    \setlength\tabcolsep{6pt}
    \caption{Effects of clean set size on Trojan mitigation task.}
    \vspace{0.1cm}
    \label{tab:ablation_mitigation_datanum}
\begin{tabular}{@{}ccc@{}}
\toprule
Samples Per Class & BA      & ASR    \\ \midrule
5                & 91.03\% & 0.08\% \\
10               & 91.79\% & 0.04\% \\
50               & 91.66\% & 0.02\% \\
100              & 91.66\% & 0.06\% \\ \bottomrule
\end{tabular}
\end{table}

\subsection{Visualization of Reverse-Engineered Trojans}\label{sec:eval_visualization}

\begin{figure}[]
    \centering
    \footnotesize
    \includegraphics[width=.8\columnwidth]{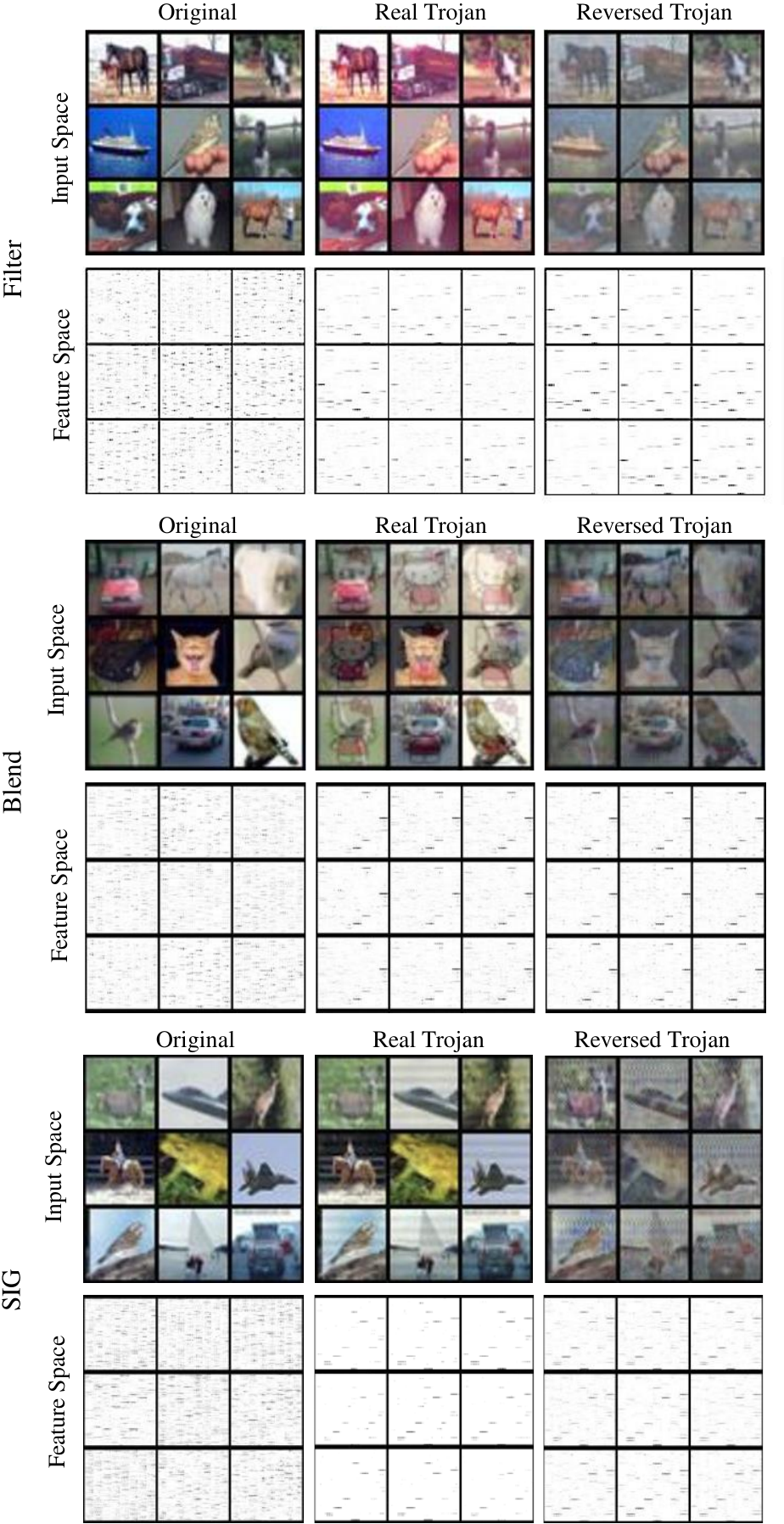}
    \caption{Visualization of the input space and the feature space for original inputs, real Trojan inputs, and reverse-engineered Trojan inputs.}\label{fig:reversed_triggers}
\end{figure}

To understand our method and study if it can reverse-engineer Trojans accurately, we visualize the inputs and inner features of clean samples, real Trojan samples, and reversed Trojan samples on nine randomly selected samples in \autoref{fig:reversed_triggers}.
The model is ResNet18 injected with Filter Trojan~\cite{liu2019abs}, Blend Trojan~\cite{chen2017targeted} and SIG Trojan~\cite{barni2019new}.
In the feature space, the reverse-engineered Trojan is close to the real Trojan, demonstrating the effectiveness of our reverse-engineering method.

\subsection{Generalization}\label{sec:generalization}

\textbf{Performance on mitigation task for more attacks.}
To measure the effectiveness of \sys on Trojan mitigation task, we use more Trojan attacks and report BA and ASR of our method.
Besides the results of BadNets~\cite{gu2017badnets}, Filter~\cite{liu2019abs}, WaNet~\cite{nguyen2021wanet} and IA~\cite{nguyen2020input} in \autoref{tab:unlearning}, in \autoref{tab:mitigation_more_attacks}, we also show the BA and ASR on LS~\cite{gu2017badnets}, CL~\cite{turner2019label} and SIG~\cite{barni2019new}.
The dataset and the model used is CIFAR-10 and ResNet18, respectively.
For LS, CL, and SIG, the ASR of \sys is 1.15\%, 2.62\%, and 1.22\%, which are 80.01, 33.18, and 81.22 times lower than that of undefended models.
As can be observed, \sys can effectively reduce the ASR while keeping the BA nearly unchanged.
Thus, \sys is robust to different attacks on mitigation task.

\begin{minipage}[htbp]{\textwidth}
    \begin{minipage}{0.5\textwidth}
        \centering
          \scriptsize
          \captionof{table}{Mitigation results for more attacks.}
          \label{tab:mitigation_more_attacks}
          \setlength\tabcolsep{3pt}
          \scalebox{1}{
            \begin{tabular}{@{}ccccccc@{}}
            \toprule
            \multirow{2}{*}{Attack} &  & \multicolumn{2}{c}{Undefended} &  & \multicolumn{2}{c}{FeatureRE} \\ \cmidrule(lr){3-4} \cmidrule(l){6-7} 
                                    &  & BA            & ASR            &  & BA          & ASR        \\ \midrule
            LS                      &  & 93.66\%       & 92.02\%        &  & 92.86\%     & 1.15\%     \\
            CL                      &  & 93.51\%       & 86.94\%        &  & 92.94\%     & 2.62\%     \\
            SIG                     &  & 93.73\%       & 99.09\%        &  & 93.47\%     & 1.22\%     \\ \bottomrule
            \end{tabular}}
            \end{minipage}
    \begin{minipage}{0.5\textwidth}
        \centering
          \scriptsize
          \captionof{table}{Detection accuracy on more models.}
          \label{tab:generalization_to_more_models}
          \setlength\tabcolsep{3pt}
          \scalebox{1}{
        \begin{tabular}{@{}cccccc@{}}
        \toprule
        Attack  & VGG16 & ResNet18 & PRN18 & LeNet & 4Conv+2FC \\ \midrule
        BadNets & 95\%  & 95\%     & 100\% & 100\% & 95\%      \\
        Filter  & 90\%  & 90\%     & 95\%  & 90\%  & 100\%     \\
        WaNet   & 90\%  & 95\%     & 90\%  & 90\%  & 95\%      \\
        IA      & 90\%  & 90\%     & 90\%  & 95\%  & 95\%      \\
        LS      & 85\%  & 90\%     & 85\%  & 80\%  & 85\%      \\
        CL      & 80\%  & 85\%     & 85\%  & 80\%  & 90\%      \\
        SIG     & 95\%  & 95\%     & 90\%  & 90\%  & 90\%      \\ \bottomrule
        \end{tabular}}
        \end{minipage}
\vspace{0.3cm}
\end{minipage}

\textbf{Generalization to different models.}
To understand the generalization of \sys to different model architectures, we evaluate its detection accuracy on BadNets~\cite{gu2017badnets}, Filter~\cite{liu2019abs}, WaNet~\cite{nguyen2021wanet}, IA~\cite{nguyen2020input}, LS~\cite{gu2017badnets}, CL~\cite{turner2019label}, and SIG~\cite{barni2019new} attacks using VGG16~\cite{simonyan2014very}, ResNet18~\cite{he2016deep}, Preact-ResNet18 (PRN18)~\cite{he2016identity}, LeNet5~\cite{lecun1998gradient}, and 4Conv+2FC~\cite{xu2019detecting}.
The results are summarized in \autoref{tab:generalization_to_more_models}.
In \autoref{tab:generalization_to_large_size}, we also report \sys's performance on a larger model (i.e., Wide-ResNet34~\cite{zagoruyko2016wide}).
In all settings, the detection accuracy is above 80\%, and the average detection accuracy on VGG16, ResNet18, and PRN18 is 89.26\%, 91.43\%, and 90.71\%, respectively.
\sys achieves high detection accuracy on all different models, demonstrating it is generalizable to different model architectures and larger models.

\textbf{Generalization to large input size.}
To see if \sys can generalize to large datasets, we report its accuracy on the ImageNette\footnote{https://github.com/fastai/imagenette} dataset under different attacks.
The input size of ImageNette is 3 \(\times \) 224 \(\times \) 224.
The model architecture used here is Wide-ResNet34~\cite{zagoruyko2016wide}.
For each attack, we have 5 Trojaned models.
We also train 5 benign models.
The results are in \autoref{tab:generalization_to_large_size}.
For all different attacks, the detection accuracy of \sys is above 80\%.
The average detection accuracy on a large input size is 91.43\%.
Thus, our method can generalize to large input sizes.

\begin{table}[H]
    \centering
    \scriptsize
    \caption{Detection accuracy on large input size.}\label{tab:generalization_to_large_size}
    \vspace{0.1cm}
    \begin{tabular}{@{}cccccc@{}}
    \toprule
    Attack  & TP & FP & FN & TN & Acc   \\ \midrule
    BadNets & 5  & 0  & 0  & 5  & 100\% \\
    Filter  & 4  & 0  & 1  & 5  & 90\%  \\
    WaNet   & 4  & 0  & 1  & 5  & 90\%  \\
    IA      & 5  & 0  & 0  & 5  & 100\% \\
    LS      & 3  & 0  & 2  & 5  & 80\%  \\
    CL      & 3  & 0  & 2  & 5  & 80\%  \\
    SIG     & 5  & 0  & 0  & 5  & 100\% \\ \bottomrule
    \end{tabular}
\end{table}

\subsection{Efficiency}\label{sec:efficiency}

In this section, we measure the efficiency of \sys.
Like existing reverse-engineering methods~\cite{wang2019neural,guo2019tabor,chen2019deepinspect}, it scans all labels.
We optimize this process with a K-arm scheduler~\cite{shen2021backdoor}, which uses the Multi-Arm Bandit to iteratively and stochastically select the most promising labels for optimization.
We measure the average runtime on the CIFAR-10 and ImageNet datasets.
The model used is ResNet18.
The running time on CIFAR-10 and ImageNet are 530.8s and 8934.5s, respectively.

\subsection{Discussions}\label{sec:findings_evaluation}
One finding we have is that using later layers to conduct the reverse-engineering is relatively better than using earlier layers (more results and details can be found in \autoref{sec:split}).
We also found that \sys's performance under the clean-label attack is relatively worse than that of other attacks.
We suspect this is because the benign and Trojan features of the clean-label attack are highly mixed.
As a consequence, the clean label attack has lower ASR than other attacks.
For example, the ASR of the clean-label attack and BadNets are 86.94\% and 100.00\%, respectively.

\end{document}